\documentstyle[12pt]{article}
\newcommand{\bea}{\begin{eqnarray}}
\newcommand{\eea}{\end{eqnarray}}
\newcommand{\ba}{\begin{array}}
\newcommand{\ea}{\end{array}}
\newcommand{\vx}{{\bar x}}
\newcommand{\vp}{{\bar p}}

\newcommand{\be}{\begin{equation}}
\newcommand{\ee}{\end{equation}}

\newcommand{\bi}{\bibitem}
\newcommand{\wvp}{\bar P}
\newcommand{\st}{\stackrel}
\newcommand{\ms}{\mathstrut}
\newcommand{\ds}{\displaystyle}
\begin{document}

\begin{titlepage}
\pagestyle{empty}
\vspace{1.0cm}
\begin{flushright}
Preprint GSI 97-72
\end{flushright}
\vspace{1.0cm}
\begin{center} \begin{Large}
{\bf
Dynamical  derivation of a quantum kinetic equation\\ for  particle
production in the Schwinger mechanism}\\ \vspace{5mm}
\end{Large}
\vskip 0.7in
{ \large S.A.
Smolyansky$^1$, G. R\"opke$^{2,3}$, S. Schmidt$^{3,4}$, D.Blaschke$^3$,\\
V.D. Toneev$^{2,5}$ and A.V. Prozorkevich$^1$ }

\vspace{5mm}

{$^1$ Physics Department, Saratov State University, Saratov, Russia}\\
{$^2$ Bogoliubov Laboratory of Theoretical Physics, \\
   Joint Institute for Nuclear Research, Dubna, Russia}\\
{$^3$ Fachbereich Physik, Universit\"at Rostock, Germany}\\
{$^4$ School of Physics and Astronomy, Tel Aviv University, Tel
Aviv, Israel}\\
{$^5$ Gesellschaft f\"ur Schwerionenforschung, Darmstadt, Germany}

\end{center}
\vspace{1.2cm}
\begin{abstract}

A quantum kinetic equation has been derived for the
description of pair production in a time-dependent homogeneous
electric field $E(t)$. As a source term, the Schwinger mechanism
for particle creation is incorporated. Possible particle
production  due to collisions and collisional damping are
neglected.  The main result is a closed kinetic equation of 
the non-Markovian type. In the low density approximation, the source
term is reduced to the leading part of the well known Schwinger
formula for the probability of pair creation. We compare the
formula obtained with other
approaches and discuss the differences. 
\end{abstract}
\end{titlepage}

\section{Introduction}
Particle production in ultrarelativistic heavy-ion collisions
raises a number of challenging problems. One of those
interesting questions is how to incorporate the mechanism of
particle creation in a dynamical kinetic
theory \cite{BC,KM,GKM}. In the framework of a flux tube model,
a lot of promising researches has been carried out. In the
scenario where a chromo-electric field is generated by a
nucleus-nucleus collision, the production of pairs can be
described by the Schwinger mechanism \cite{Sau,HE,Sch}.
The charged particles produced generate a
field which, in its turn, influences  the initial electric
field and may cause plasma oscillations. Recently the
interesting question of this back reaction  has been
analyzed within a field theoretical approach
\cite{CM,KESCM1,KESCM2}. The results of a simple
phenomenological consideration based on kinetic equations
and the full field-theoretical treatment
\cite{CEKMS,KES,CHKMPA,Eis1} agree with each other.
The source term which occurs in such a modified Boltzmann
equation was derived phenomenologically in \cite{BE}.
However, the appearence of this source term in
relativistic transport theory which may be of non-Markovian
character \cite{Rau,RM} is not yet fully  understood.

In the present work, a kinetic equation is derived in a
dynamically consistent way for the time evolution of  the pair
creation in a time-dependent and spatially homogeneous electric
field, a process referred often to as
the Schwinger mechanism. This derivation is based on the
Bogoliubov transformation for field operators between
in-state and instantaneous states. In contrast
with phenomenological approaches, the source term of the particle
production is of non-Markovian character. The kinetic equation
derived reproduces the Schwinger result in the asymptotic  limit
and in the low density approximation.

The paper is organized as follows. In Section 2 we sketch some
important features of our approach to be used in Section 3
for deriving the basic kinetic equation describing particle creation in
strong electric fields.  Finally we discuss the properties of the
equation obtained.

\normalsize
\section{Model description}
We use a simple  field-theoretical model  to treat fermions
in an external electric field.
To be described by the vector potential $A^\mu=(0,0,0,A^3(t))$
\footnote{We use the units $\hbar = c = 1$ and the
metric is chosen to be $g^{\mu\nu} = {\rm diag}(1,-1,-1,-1)$.}
the electric field $E^3(t)$  is assumed to be time-dependent but
homogeneous in space and hence
$E^1 =E^2=0$, 
\bea\label{200}
E(t)=E^3(t)=-{\dot A}\ms^3(t)=-dA^3(t)/dt ~. 
\eea
This quasi-classical electric field interacts with a spinor field $\psi$ of
fermions. We look for solutions of the Dirac equation where eigenstates
with the  positive and negative frequencies are  represented  in the form
\footnote{It is convenient to work in the discrete momentum
representation assuming the limited volume of the system, $V=L^3$.}
\cite{NN,PM,P,MP,CMM}:
\bea\label{10} \psi^{(\pm)}_{\vp r}(x)& =& L^{-3/2}
\bigg[i\gamma^0\partial_0+\gamma^kp_k-e\gamma^3A_3(t)+m\bigg] \
\chi^{(\pm)}(\vp,t) \ R_r \ {\rm e}^{i\vp\bar x},
\eea
where $k=1,2,3$ and $R_r$ $(r = 1,2)$ is an
eigenvector of the matrix $\gamma^0\gamma^3$
\be\label{30}
R_1 =\left(\begin{array}{r}0\\1\\0\\-1\end{array}\right)\,\,,\hspace{4cm}
R_2 =\left(\begin{array}{r}1\\0\\-1\\0\end{array}\right)\,\,\, ,
\ee
so that $R^+_r R_s = 2\delta _{rs} \ .$
The functions $\chi^{(\pm)}(\vp,t)$ are related to the
oscillator-type equation
\bea
\label{40} {\ddot
\chi}^{(\pm)}(\vp,t)&=&-\bigg(\omega^2(\vp,t)+ie{\dot
A}_3(t)\bigg) \ \chi^{(\pm)}(\vp,t)\,\,,
\eea
with $\omega^2(\vp,t)=\mu^2+P^2_3$,
$\mu^2=m^2+p^2_\perp$ being the transverse mass and
$P_3=p_3-eA_3(t)$. The solutions $\chi^{(\pm)}(\vp,t)$ of
Eq.(\ref{40}) for positive and negative frequencies are
defined by their asymptotic behavior at $t\rightarrow -\infty$
\be\label{60}
\chi^{(\pm)}(\vp,t) \lefteqn{\phantom{tt}\sim}\ba{c}\\{\scriptstyle t\to 
- \infty} \ea \exp{\big(\pm i\omega_-(\vp)\,t\big)}\,\,,
\ee
where $\omega_-(\vp)=\lim\limits_{t\to-\infty}\omega(\vp,t)$.
In the following we will sometimes drop  the momentum dependence
of the frequency and use the short notation $\omega(\vp,t)=\omega(t)$.
 Note that the system of the spinor functions
 (\ref{10}) is complete and orthonormalized. So,
the field operators $\psi(x)$ and ${\bar \psi}(x)$ can be
decomposed in the spinor functions (\ref{10})  as follows:
\bea\label{70} \ba{lcl}
\psi(x)&=&\sum\limits_{r,\vp}
\bigg[\psi^{(-)}_{\vp r}(x) \ a^{(-)}_{\vp r}+\psi^{(+)}_{\vp r}(x) \
a^{(+)}_{-\vp r}\bigg] \,\,,\\  \\
{\bar\psi(x)}&=&\sum\limits_{r,\vp}\bigg[{\bar \psi}^{(-)}_{\vp r}(x) \
{\st{*}{a}}\ms^{(+)}_{\vp r}+{\bar \psi}^{(+)}_{\vp r}(x) \
{\st{*}{a}}\ms^{(-)}_{-\vp r}\bigg]\,\,. \ea \eea
The operators
${\st{*}{a}}\ms^{(\pm)}_{\vp r}$ and $a^{(\pm)}_{\vp r}$ describe 
\footnote{Below we follow the notation of Ref. \cite{bs}.}
the creation and annihilation of electrons and positrons
in the in-state at $t\!\to\!-\infty$ with the vacuum state
$|0_{in}\!>$. The evolution will affect the vacuum state and mix
the positive and negative energies resulting in non-diagonal
terms of eigenstates responsible for pair creation. The
diagonalization of the Hamiltonian 
for a Dirac-particle in the homogeneous electric field (\ref{200})
is achieved by means of
the time-dependent Bogoliubov  transformation
\bea\label{90}\ba{lclcl} b^{(-)}_{\vp r}(t)
&=&\alpha_\vp (t)\ a^{(-)}_{\vp r}&+&
\beta_\vp (t)\ a^{(+)}_{-\vp r}\ ,
\\ \\ {\st{*}{b}}\ms^{(-)}_{\vp r}(t)&=&\alpha_{-\vp} (t) \
{\st{*}{a}}\ms^{(-)}_{\vp r}&-&\beta_{-\vp}(t) \
{\st{*}{a}}\ms^{(+)}_{-\vp r}
\ea \eea
with the condition
\be\label{110}
|\alpha_\vp(t)|^2+|\beta_\vp(t)|^2=1\,\,.  \ee
Here, the operators
${\st{*}{b}}\ms^{(\pm)}_{\vp r}$ and $b^{(\pm)}_{\vp r}$
 describe the creation and annihilation of quasiparticles
 at the time $t$ with the instantaneous
vacuum $|0_t\!>$. The substitution of
Eqs.(\ref{90}) into Eqs.(\ref{70}) leads to the new
representation of the field operators, e.g.
\bea\label{120}
\psi(x)&=&\sum_{r,\vp}\bigg[\Psi^{(-)}_{\vp r}(x) \
b^{(-)}_{\vp r}(t) + \Psi^{(+)}_{\vp r}(x) \ b^{(+)}_{-\vp
r}(t)\bigg]\,\,.
\eea
The new basis functions $\Psi^{(\pm)}_{\vp r}(x)$ are
connected with the former ones (\ref{10}) by a canonical
transformation like Eqs.(\ref{90})
\bea
\label{130}\ba{lcl}
\psi^{(-)}_{\vp r}(x)&=&\alpha_\vp(t) \ \Psi^{(-)}_{\vp r} -
\beta^*_\vp(t) \ \Psi^{(+)}_{\vp r}\,\,,\\ \\
\psi^{(+)}_{\vp r}(x)&=&\alpha^*_\vp(t) \ \Psi^{(+)}
_{\vp r} + \beta_\vp(t) \ \Psi^{(-)}_{\vp r}\,\,.
\ea \eea
Therefore one may assume that the functions $\Psi^{(\pm)}_
{\vp r}$ have a spin structure  similar to $\psi^{(+)}_{\vp r}$
given by Eq.(\ref{10}),
\be
\label{150}
\Psi^{(\pm)}_{\vp r}(x)=  L^{-3/2}
\bigg[i\gamma^0\partial_0+\gamma^kp_k-e\gamma^3A_3(t)
+m\bigg]\phi^{(\pm)}_{\vp}(x) \ R_r \
{\rm e}^{\pm i\Theta(t)} {\rm e}^{i\vp\vx},
\ee
where the dynamical phase is defined as
\be\label{160}
\Theta(t) = \int^t_{t_0}dt'\omega(t')\,\,
\ee
and $\phi^{(\pm)}_{\vp}$ are yet unknown functions. Substitution
of (\ref{150}) into Eqs.(\ref{130}) leads to
the relations
\bea\label{170}\ba{lclcl}
\chi ^{(-)}(\vp,t)&=&\alpha_\vp(t) \ \phi_\vp^{(-)}(t) \ {\rm
e}^{-i\Theta(t)}&-&\beta^*_\vp(t) \ \phi^{(+)}_\vp(t) \ {\rm
e}^{i\Theta(t)}\,\,,\\ \\
\chi^{(+)}(\vp,t)&=&\alpha^*_\vp(t) \ \phi_\vp^{(+)}(t) \ {\rm
e}^{i\Theta(t)}&+&\beta_\vp(t) \ \phi^{(-)}_\vp(t) \
{\rm e}^{-i\Theta(t)}\,\,.
\ea \eea
Taking into account that the functions $\chi ^{(\pm)}(\vp,t)$
are defined by Eq.(\ref{40}), we can find the equations for
functions $\alpha_\vp(t)$ and $\beta_\vp(t)$. According to the
Lagrange method it is possible to introduce
additional conditions to Eqs.(\ref{170})
\bea\label{190}\ba{lclcl}
 {\dot \chi}^{(-)}(\vp,t)&=&-i\omega(t)\ \bigg[\alpha_\vp(t) \
\phi_\vp^{(-)}(t) \ {\rm e}^{-i \Theta(t)}&+& \beta^*_\vp(t) \
\phi^{(+)}_\vp(t) \ {\rm e}^{i\Theta(t)}\,\,\bigg]\, \,,\\ \\{\dot
	\chi}^{(+)}(\vp,t)&=&\quad i\omega(t)\ \bigg[  \alpha^*_\vp(t) \
\phi_\vp^{(+)}(t) \ {\rm e}^{i\Theta(t)}&-& \beta_\vp(t) \
\phi^{(-)}_\vp (t) \ {\rm e}^{-i\Theta(t)}\bigg]\,\,.
\ea \eea
Let us differentiate these equations repeatedly and make use
of Eqs.(\ref{40}) and (\ref{170}). Then, choosing  the
functions $\phi^{(\pm)}_\vp(t)$ in the form
\be\label{210}
\phi^{(\pm)}_\vp(t)=\sqrt{\frac{\omega(t)\pm P_3}{\omega(t)}} \ ,
\ee
we obtain the following equations \cite{PM,P,MP,CMM}:
\bea\label{240}\ba{lcl}  {\dot
\alpha}_\vp(t)&=&\quad {\ds\frac{eE(t)\mu}{2\omega^2(t)}}\
\beta^*_\vp(t) \
{\rm e}^{2i\Theta(t)}\,\,,\\ \\ {\dot
\beta}^*_\vp(t)&=&-{\ds\frac{eE(t)\mu}{2\omega^2(t)}}\
\alpha_\vp(t) \ {\rm
e}^{-2i\Theta(t)}\,\,.
\ea \eea
One should note also the usefull relation for these
transformation coefficients
\bea\label{220}
\ba{lcl} \alpha_\vp(t)&=&{\ds\frac{1}{2\sqrt{\omega(t) \
(\omega(t)-P_3)}}}\bigg(\omega(t)\ \chi^{(-)}(\vp,t)+i \ {\dot
\chi}^{(-)}(\vp,t)\bigg) \ {\rm e}^{i\Theta(t)}\,\,,\\  \\
\beta^*_\vp(t)&=&-{\ds\frac{1}{2\sqrt{\omega(t) \
(\omega(t)-P_3)}}}\bigg(\omega(t) \ \chi^{(-)}(\vp,t)-i \ {\dot
 \chi}^{(-)}(\vp,t)\bigg) \ {\rm e}^{-i\Theta(t)}\,\,.
\ea \eea
It is convenient to introduce  new operators
\bea\label{260} \ba{lcl}
c^{(\pm)}_{\vp r}(t)=b^{(\pm)}_{\vp r}(t) \
\exp{\big(\pm i\Theta(t)\big)},\\ \\
{\st{*}{c}}\ms^{(\pm)}_{\vp r}(t)=
{\st{*}{b}}\ms^{(\pm)}_{\vp r}(t) \exp {\big(\pm
i\Theta(t)\big)} \ .
\ea \eea
It is easy to show (e.g., see \cite{bs}) that these operators
are consistent with the rules of the
Hermitian conjugation for creation and annihilation operators, i.e.
\bea\label{265}
[c^{(\pm)}_{\vp r}(t)]^+ ={\st{*}{c}}\ms^{(\mp)}_{\vp r}(t)
\eea
and satisfy the Heisenberg-like  equations of motion
\bea\label{270} \ba{lcl}
{\ds\frac{dc^{(\pm)}_{\vp r}(t)}{dt}=\pm\frac{e E(t)\mu}{2\omega^2(t)}} \
c^{(\mp)}_{-\vp r}(t) + i \ [H(t), \ c^{(\pm)}_{\vp r}(t)]\,\,,  \\ \\
{\ds\frac{d{\st{*}{c}}\ms^{(\pm)}_{\vp r}(t)}{dt}=\mp\frac{e E(t)\mu}
{2\omega^2(t)}}\ {\st{*}{c}}\ms^{(\mp)}_{-\vp r}(t)+
i \ [H(t),\ \st{*}{c}\ms^{(\pm)}_{\vp r}(t)] \ ,
\ea \eea
where $H(t)$ is the Hamiltonian of the quasiparticle system
\be\label{280}
H(t)=\sum_{r,\vp} \omega(t)\bigg({\st{*}{c}}\ms^{(+)}_{\vp r}(t) \
c^{(-)}_{\vp r}(t)-{\st{*}{c}}\ms^{(-)}_{-\vp r}(t) \ c^{(+)}_{-\vp
r}(t)\bigg)\,\, .
\ee
The first term on the r.h.s of Eqs.(\ref{270}) is caused by
the unitary non-equivalence of the in-representation and the
quasiparticle one.
  The commutation relations
\be \label{290}
[{\st{*}{c}}\ms^{(-)}_{\vp r}(t), \ c^{(+)}_{\vp'
r'}(t)]_+=[c^{(-)}_{\vp r}(t), \ {\st{*}{c}}\ms^{(+)}_{\vp'
r'}(t)]_+=\delta_{rr'} \ \delta_{\vp \vp'}
\ee
are fulfilled for the operators (\ref{260}) .

\section{ Dynamics of pair creation  for fermions}
Now we shall consider the kinetic equation for the distribution
function of electrons with the momentum $\vp$ and  spin $r$
\be\label{300}
f_r(\vp,t) = <0_{in}|{\st{*}{b}}\ms^{(+)}_{\vp r}(t) \
b^{(-)}_{\vp r}(t)|0_{in}> =
<0_{in}|{\st{*}{c}}\ms^{(+)}_{\vp r}(t) \
c^{(-)}_{\vp r}(t)|0_{in}>\,\,.
\ee
According to the charge conservation the distribution functions
for electrons
and positrons are equal $f_r(\vp,t) = {\bar f}_r(\vp,t)$, where
\be \label{320}
{\bar f}_r(\vp,t) =<0_{in}|b^{(+)}_{-\vp r}(t) \
{\st{*}{b}}\ms^{(-)}_{-\vp r}(t)|0_{in}>=<
0_{in}|c^{(+)}_{-\vp r}(t) \
{\st{*}{c}}\ms^{(-)}_{-\vp r}(t)|0_{in}>\,\,.
\ee

The distribution functions (\ref{300}) and (\ref{320}) are normalized to
the total particle number $N(t)$ of the system at the given time moment,
\be\label{330} \sum \limits_{r,\vp}f_r(\vp ,t)= \sum \limits_{r,\vp}
\bar f_r(\vp ,t)=N(t)\ . \ee

By the time differentiation of Eq. (\ref{300}) we get
\be\label{340}
\frac{d f_r(\vp,t)}{dt}= -\frac{e\mu E(t)}{2\omega^2(t)} \
\bigg[\Phi_r^{(+)}
(\vp,t)+\Phi_r^{(-)}(\vp,t)\bigg]\,\,.
\ee
We have used here the equation of motion (\ref{270}) and evaluated
the occurring commutator. The functions $\Phi_r^{(\pm)}$ in
Eq.(\ref{340}) describe the creation ($\Phi_r^{(+)}$) and
annihilation  ($\Phi_r^{(-)}$) of an electron-positron pair
in the external electric field $E(t)$. They are given as
\bea \label{350} \ba{lcl}
\Phi_r^{(-)}(\vp,t)&=& <0_{in}|{\st{*}{c}}\ms^{(-)}_{-\vp r}(t) \
c^{(-)}_{\vp r}(t)|0_{in}>\,\,,\\ \\ \Phi_r^{(+)}(\vp,t)&=&
<0_{in}|{\st{*}{c}}\ms^{(+)}_{\vp r}(t) \ c^{(+)}_{-\vp r}(t)|0_{in}>\,\,.
\ea \eea
It is straightforward to evaluate the derivatives of these functions. 
Then, by applying
the equations of motion (\ref{270}), we get
\be\label{360}
\frac{d\Phi_r^{(\pm)}(\vp,t)}{dt}=\frac{e\mu
E(t)}{2\omega^2(t)}\bigg[2f_r(\vp,t)-1\bigg]\pm2i\omega(t) \
\Phi_r^{(\pm)} (\vp,t)\,\,,
\ee
where due to charge  neutrality of the system, the relation
$f_r(\vp,t) = {\bar f}_r(\vp,t)$ is used. The solution of
 Eq.(\ref{360}) may be written in the following integral form:
\be\label{370}
\Phi_r^{(\pm)}(\vp,t) = \frac{\mu}{2}\int_{-\infty}^t dt'\frac{e
E(t')}{\omega^2(t')}\bigg[2f_r(\vp,t')-1\bigg]{\rm e}^{\pm
2i[\Theta(t)-\Theta(t')]}\,\,.
\ee
Here we proceeded to the limit $t_0\to -\infty$
for  functions $\Theta (t)$ and $\Theta (t')$ in r.h.s. of
Eq.(\ref{370}) (see the definition (\ref{160}))
assuming  that $\lim\limits_{t\to -\infty}A^3(t)=A^3_-=0$, i.e.
the fields $\Phi^{(\pm)}_r(\vp ,t)$ vanish at $t\to-\infty$.
Inserting Eq.(\ref{370}) into the r.h.s of Eq.(\ref{340})
we obtain
\be\label{380} \frac{df_r(\vp,t)}{dt}=
\frac{e\mu
E(t)}{2\omega^2(t)}\int_{-\infty}^t dt' \frac{e\mu
E(t')}{\omega^2(t')}\bigg[1-2f_r(\vp,t')\bigg]\cos
\bigg(2[\Theta(t)-\Theta(t')]\bigg)\,\,.
\ee
It is seen from (\ref{380}), that the distribution function does
not depend on spin, so $f_r = f$.

Now let the volume of the system goes to infinity, $V\rightarrow
 \infty$. To do this limiting transition, we use the rule
$$ {\left(\frac{2\pi}{L}\right)}^3\sum\limits_\vp ...\to
\int d^3p ... \ , $$ take into account the normalization conditions
(\ref{330}) and introduce the distribution function $F(\vp,t)$ normalized
to the particles number density, i.e.
\bea
(2\pi)^{-3}g\int d^3 p \ F(\vp,t)=n(t).
\label{390}
\eea
where $g=2$ is the degeneracy factor for electrons.
Then, with the substitution $f(\vp,t)\to F(\vp,t)$, 
and after replacing the canonical momentum $\vp$ by the
kinetic one, $p\to P=p-eA$, where the 3-momentum is now defined as
$\wvp (p^1, p^2, P^3)$,
the kinetic equation (\ref{380}) is reduced to the final form:
\be
\label{400}
\frac{dF(\wvp,t)}{dt}=\frac{\partial F(\wvp,t)
}{\partial t}+eE(t)\frac{\partial F(\wvp,t) }{\partial P_3}=
{\cal S}(\wvp,t)\,\,,
\ee
with the  Schwinger source term
\be
\label{410}
{\cal S}(\wvp,t) = \frac{e\mu
E(t)}{2\omega^2(t)}\int_{-\infty}^t dt' \frac{e\mu
E(t')}{\omega^2(t')}\bigg[1-2F(\wvp,t')\bigg]\cos
\bigg(2[\Theta(t)-\Theta(t')]\bigg)\,\,.
\ee
Recently, a kinetic
equation similar to (\ref{400}) has been derived within a projection 
operator formalism for the case of a time-independent
electric field in Ref. \cite{Rau} where it was first noted that this source 
term has non-Markovian character.  
The presence of the Pauli blocking factor $[1-2F(\wvp,t)]$ in the source term
has been obtained earlier in Ref. \cite{KESCM2}.
We would like to emphasize the closed form of the kinetic equation in the
present work where the source term does not include the anomalous
distribution functions (\ref{350}) for fermion-antifermion pair creation 
(annihilation).

\section{Concluding remarks }
Based on microscopic dynamics,  the closed kinetic equation
(\ref{400}) has been derived
for fermion evolution in a time-dependent, homogeneous
electric field with the Schwinger source term (\ref{410})
to be characterized by the following features:
The kinetic equation (\ref{400}) is of non-Markovian type
 due to the explicit dependence of the source term on the
whole pre-history accounting for a memory effect. The 
difference of dynamical phases, $\Theta (t)-\Theta(t')$, under
the integral (\ref{380})  generates high frequency oscillations.
The appearence of such a source term violates the time reversal
symmetry and may lead to entropy production due to the pair
creation. As was noted in Refs. \cite{Rau,RM}, this may result also
in oscillations of the relevant entropy and thus in temporaryn
(on the scale of the memory time) violations of the H-theorem.
The phase space occupation of the created fermions is taken into
account by the Pauli blocking factor $1-2F(\wvp,t')$ in (\ref{410}).

It is worthwhile to compare our result with the equation obtained
phenomenologically in Refs. \cite{KESCM2,CEKMS}
\be
\label{420}
{\cal S}^{{\rm ph}}(\wvp,t)= - [1-2F(\wvp,t)] 
\ |eE(t)| \ \ln\bigg[1-\exp\bigg(-\frac{\pi\mu^2}
{|eE(t)|}\bigg)\bigg] \ \delta(P_3)\,\,.
\ee
By comparing Eqs.(\ref{410}) and (\ref{420}) we see that the
essential difference between them is  in the non-Markovian
character of Eq.(\ref{410}). This rather intriguing feature
leads to  an accumulation of created particles on
the background of rapid oscillations.

Finally let us check the semi-classical limit for the case of
a constant electric field, $E(t) = E$. From Eq.(\ref{410})
at $t\to +\infty$ and
in the low density limit $F(\wvp,t)\ll 1$, the pair production
rate is reduced to the following formula:
\be
\label{430}
{\cal S}^{{\rm cl}}=\lim_{t\to +\infty}(2\pi)^{-3}g\int
d^3P \
{\cal S}(\wvp,t)=\frac{e^2E^2}{4\pi^3}\exp
\bigg(-\frac{\pi m^2}{|eE|}\bigg)
\ee
which covers the well-known Schwinger result
obtained in the semi-classical
approximation.
It is of interest to note that in contrast with this
semi-classical limit the source term (\ref{410}) is not
positively defined, in general.

Our consideration can also be applied to scalar systems
described by the Klein-Gordon equation. Under the same
assumptions we obtain Eq. (\ref{400})  (see Appendix A )
with the source term
\be
\label{440}
{\cal S}(\wvp,t) = \frac{e\mu E(t)}{2\omega^2(t)}
\int_{-\infty}^t dt' \frac{e\mu
E(t')}{\omega^2(t')} \ \bigg[1+2F(\wvp,t')\bigg]\cos
\bigg(2[\Theta(t)-\Theta(t')]\bigg)\,\,.
\ee
The difference between  the sources (\ref{410}) and (\ref{440}) is
only in the  phase space occupation factors  due to the
different quantum statistics of particles involved: the Pauli blocking  
factor $1- 2F(\wvp,t)$ which is present in the source term for the creation 
of a fermion pair has to be replaced by a Bose enhancement factor 
$1+ 2F(\wvp,t)$ when a pair of bosons is created (stimulated pair production).

Besides of the numerical solution of the derived kinetic equation,
in our future studies we would like to incorporate collisions
in  this formalism and apply our method for
the back reaction problem. In this case  the time-dependent
electric field must be defined by means  of the Maxwell's
equation with the
spatially homogeneous current $j(t)=-(1/4\pi )\dot E (t)$.

\section{Acknowledgements}
The authors wish to thank V.G.  Morozov for interest in this work,
 M.K.  Volkov for discussions and J. Eisenberg, J. Rau
 and   N.B.  Perkins for useful  comments.
The authors gratefully acknowledge the hospitality at the University
of Rostock (S.A.S.), the JINR Dubna (D.B. and S.S.) and the GSI Darmstadt 
(V.D.T.)  where part of this work has been carried out. 
This work was supported in part by the
Russian State Committee  of High-School Education
(under grant No. 95-0-6/1-53), the Minerva foundation,
the Heisenberg-Landau program and the WTZ program of the BMBF.

\appendix
\section*{Appendix A}
Here  some details are given to  derive of the kinetic equation with
the source (\ref{440}) describing the vacuum creation of scalar
mesons in a strong electric field.

The solution of the Klein-Gordon equation in the presence of the electric
field defined by the vector potential $A^\mu =(0,0,0, A^3(t))$
is taken in the form \cite {NN,PM,P,MP,CMM}
$$\phi^{(\pm)}_{\vp}(x)=L^{-3/2}[2\omega_-(p)]^{-1/2}\ e^{i\bar x\vp}
g^{(\pm)}(\vp ,t)\ ,  \eqno(A1) $$
where the functions $g^{(\pm)}(\vp ,t)$  satisfy  the
oscillator-type equation with a variable frequency $$
\ddot g^{(\pm)}(\vp ,t)+\omega^2 (\vp,t) \
g^{(\pm)}(\vp ,t)=0\ .\eqno(A2)$$
Solutions of Eq.(A2) for positive and negative frequencies are defined
by their asymptotic behaviour at $t\to -\infty$ similarly to
Eq.(\ref{60}). The mesonic functions
(A1) are normalized  (in the discrete momentum
representation) in the following way: $$
i\int\limits_V d^3x \phi^{(\pm)*}_{\vp}(x)\st{\leftrightarrow}{\partial _0}
\phi^{(\pm)}_{\vp '}(x)=\mp \delta_{\vp\vp'}\ . \eqno(A3) $$

The field operator in the in-state is defined as $$
\phi (x)=\int d^3p \ [\phi^{(+)}_{-\vp}(x) \ a^{(+)}_{\vp} +
\phi^{(-)}_{\vp}(x) \ a^{(-)}_{\vp}]\ , \eqno(A4) $$
being Hermitian conjugated, i.e., $
[a^{(\pm)}_{\vp }]^+ ={\st{*}{a}}\ms^{(\mp)}_{\vp }$.
The diagonalization of the Hamiltonian is achieved by the transition to
quasiparticle representation. The Bogoliubov transformation for
creation and annihilation operators of quasiparticles
has the following form:
$$\ba{lclcl}
b^{(-)}_{\vp }(t)
&=&\alpha_\vp (t)\ a^{(-)}_{\vp } &+&\beta_\vp (t)\ a^{(+)}_{-\vp }\ ,
\\ \\ b^{(+)}_{-\vp }(t)&=&\alpha^*_{\vp} (t)a^{(+)}_{-\vp }&+&
\beta^*_\vp (t)\ a^{(-)}_{\vp }\  \ea \eqno(A5) $$
with the condition $$
|\alpha_\vp(t)|^2-|\beta_\vp(t)|^2=1\,\,.  \eqno(A6) $$
The field operator (A4) may be rewritten now in terms of the
quasiparticle operators for the creation
and annihilation processes $$
\phi (x)=\int d^3p [\Phi^{(-)}_\vp (x)b^{(-)}_\vp  +\Phi^{(+)}_\vp(x)
b^{(+)}_\vp ]\ ,  \eqno(A7)    $$
where the new amplitudes $\Phi^{(\pm)}_\vp (x)$ are related to the functions
(A1) by a canonical transformation similar to (A5)
$$ \ba{lclcl}
\phi^{(-)}_{\vp}(x)&=&\alpha_\vp (t)\ \Phi^{(-)}_{\vp}(x) &+&
\beta^*_\vp(t)\ \Phi^{(+)}_{\vp}(x)
\ , \\ \\ \phi^{(+)}_{\vp}(x)&=&\alpha^*_{\vp}(t)\Phi^{(+)}_{\vp}(x)&+&
\beta_\vp (t)\ \Phi^{(-)}_{\vp}(x)\ .  \ea \eqno(A8) $$
These transformations should be considered together  with Eq.(A5).

Let us find now the equation of motion for coefficients of the Bogoliubov
transformation. It is assumed by analogy with Eq.(A1) that
$$
\Phi^{(\pm)}_{\vp}(x) =L^{-3/2} \ Q^{(\pm)}(\vp ,t) \
e^{\pm i\Theta (t)} \ e^{i\vp \bar x} \ ,
\eqno(A9) $$
where $Q^{(\pm)}(\vp ,t)$ are new unknown functions and the
dynamical phase $\Theta (t)$ is  defined by Eq.(\ref{160}).
After the  substitution of Eq.(A9) into Eq.(A8) we have
 $$ \ba{lclcl}
g^{(+)}(\vp ,t)&=&\alpha^*_\vp (t)\ Q^{(+)}(\vp ,t) \
e^{i\Theta (t)}&+&
\beta_\vp(t)\  Q^{(-)}(\vp ,t) \ e^{-i\Theta (t)}~,\\ \\
g^{(-)}(\vp ,t)&=& \alpha_\vp (t)\ Q^{(-)}(\vp ,t) \
e^{-i\Theta (t)}&+&
\beta^*_\vp(t)\  Q^{(+)}(\vp ,t) \ e^{i\Theta (t)} \ ,
\ea \eqno(A10)$$
where functions $g^{(\pm)}(\vp ,t)$  should obey  Eqs.(A2). According to
the Lagrange method (e.g. \cite{mu}), this requirement may be
associated with the following additional relations
$$ \ba{lclcl}
\dot g^{(+)}(\vp ,t)&=& i\omega(t) \  \bigg[ \ \
\alpha^*_\vp (t)\ Q^{(+)}(\vp ,t) \
e^{i\Theta (t)}&-&\beta_\vp(t) \
Q^{(-)}(\vp ,t) \
e^{-i\Theta (t)}\bigg]~, \\ \\ \dot g^{(-)}(\vp ,t)
&=&i\omega(t)\ \bigg[-\alpha_\vp\
(t)\ Q^{(-)}(\vp ,t) \ e^{-i\Theta (t)}
&+&\beta^*_\vp(t) \ Q^{(+)}(\vp ,t)  \ e^{i\Theta (t)}\ \ \bigg] \ .
\ea \eqno(A11)$$
After differentiation of Eqs. (A11) and the use of the
relations (A10) and (A2) we obtain the equations of motion for the
coefficients of the canonical transformation (A5):
$$
\dot\alpha_\vp (t)=\frac{\dot\omega(t)}{2\omega(t)} \
\beta^*_\vp(t) \
e^{2i\Theta (t)}\ ,\quad \dot\beta_\vp(t)=
\frac{\dot\omega(t)}{2\omega(t)} \
\alpha^*_\vp (t) \ e^{2i\Theta (t)}
\eqno(A12) $$
 and the relation $Q^{(\pm)}(\vp ,t)=[2\omega(t)]^{-1/2}$.
Now, introducing new creation and annihilation operators by
$$
 c^{(\pm)}_{\vp}(t)=b^{(\pm)}_{\vp}(t) \
\exp{[\pm i\Theta(t)]}\ ,      \eqno(A14) 
$$
and  using the Eqs. (A5) and (A12) we get the Heisenberg-like
equation of motion \cite{PM,P,MP,CMM}
$$
{\ds\frac{dc^{(\pm)}_{\vp}(t)}{dt}=
\frac{\dot\omega}{2\omega(t)}} \
c^{(\mp)}_{-\vp}(t) + i \ [H(t), \ c^{(\pm)}_{\vp }(t)]\,\, ,
\eqno(A14)$$
with the Hamiltonian of a quasiparticle system, $$
H(t)=\sum_{\vp} \omega(t) \ \{ {\st{*}{c}}\ms^{(+)}_{\vp}(t)\
c^{(-)}_{\vp}(t)+{\st{*}{c}}\ms^{(-)}_{-\vp }(t) \ c^{(+)}_{-\vp}
(t)\}\,\,.  \eqno(A15) $$
The  canonical transformation (A5) does not change the
form of the commutation relations
$$
[{\st{*}{c}}\ms^{(-)}_{\vp }(t), \ c^{(+)}_{\vp'}(t)]_-
=[c^{(-)}_{\vp}(t),
 \ {\st{*}{c}}\ms^{(+)}_{\vp'}(t)]_-=\delta_{\vp \vp'} .
\eqno(A16)$$

Combined with Eqs.(A15) and (A16), the equations of  motion
(A14) are sufficient for deriving the kinetic equation
(\ref{400}) with the source term (\ref{440}). The derivation
procedure is very similar to the case of the electron-positron
system  considered in Sect.3.


\end{document}